\documentclass[12pt]{article}
\usepackage{amssymb}
\usepackage{amstext}
\newcommand*\de{\mathrm{d}}
\newcommand*\De{\mathrm{D}}
\renewcommand*\epsilon{\varepsilon}
\renewcommand*\phi{\varphi}
\renewcommand*\theta{\vartheta}
\begin{document}
\title{\bf Hermitian Dirac Hamiltonian in time dependent gravitational field} 


\author{M. Leclerc \\ \small Section of Astrophysics and Astronomy, 
Department of Physics, \\ \small University of Athens, Greece}  
\date{November 14, 2005}
\maketitle 

\begin{abstract}
It is shown by a straightforward argument that the Hamiltonian 
generating the time evolution of the Dirac wave function 
in relativistic quantum mechanics 
is not hermitian  
with respect to the covariantly defined inner product whenever the background 
metric is time dependent. An alternative, hermitian,  
Hamiltonian is found and is shown to be directly related to the canonical 
field Hamiltonian used in quantum field theory. 
\end{abstract}


\section{Hermiticity of the Dirac Hamiltonian}
We start with the Dirac Lagrangian in a gravitational background 
(see \cite{hehl}) 
\begin{equation}
L = \frac{i}{2}(\bar \psi \gamma^i \De_i \psi - \De_i \bar \psi \gamma^i 
 \psi - m \bar \psi \psi),  
\end{equation}
with $\De_i \psi = \partial_i \psi - \frac{i}{4} \Gamma^{ab}_{\ \ i} 
\sigma_{ab}\psi $ and $\gamma^i = e^i_a \gamma^a$, where $\gamma^a$ 
are the constant Dirac matrices and $\sigma_{ab}$ the Lorentz generators. 
Our notations are identical 
to those used in \cite{leclerc}. 
Especially, we use latin indices from the begin of the alphabet
($a,b,c \dots $) for 
tangent space quantities and from the middle ($i,j,k \dots $)  
for curved spacetime quantities. 
The  Lorentz  connection
$\Gamma^{ab}_{\ \ i}$ ($\Gamma^{ab}_{\ \ i} = - \Gamma^{ba}_{\ \ i}$)
and the tetrad  $e^a_i$  
transform under a local Lorentz transformation  
 $\epsilon^{ab}$ as 
$ \delta \Gamma^{ab}_{\ \ i}  = - 
 \partial_i \epsilon^{ab}
- \Gamma^a_{\ ci}  \epsilon^{cb} - \Gamma^b_{\ ci}  \epsilon^{ac}, 
 \  \delta e^a_i =   \epsilon^a_{\ b} e^b_i$.  
The Lagrangian (1) is invariant under such transformations 
if $\psi $ undergoes the gauge transformation  
$ \psi \rightarrow \exp{[(-i/4) \epsilon^{ab} \sigma_{ab}}]\psi $.
The Dirac equation is derived in the form 
\begin{equation}
i \gamma^i \nabla_i \psi = m \psi,   
\end{equation}
where $ \nabla_i \psi = \partial_i \psi - \frac{i}{4} \tilde 
\Gamma^{ab}_{\ \ i}\sigma_{ab} \psi $. 
The connection $\tilde \Gamma^{ab}_{\ \ i}$ is the connection that 
effectively couples to the spinor field. If we work in the framework 
of general relativity, then $\Gamma^{ab}_{\ \ i} = \Gamma^{ab}_{\ \
  i}(e)$, i.e., the connection is not an independent field and can 
be expressed in terms of the tetrad and its derivatives. In that case, 
$\tilde \Gamma^{ab}_{\ \ i} = \Gamma^{ab}_{\ \ i} $ (and $\De_i = \nabla_i$). 
If the connection is considered to be an independent field (Poincar\'e gauge 
theory), then we have $\Gamma^{ab}_{\ \ i} = \Gamma^{ab}_{\ \ i}(e) 
+ K^{ab}_{\ \ i}$, where $K^{ab}_{\ \ i}$ is the contortion tensor. 
However, only the totally antisymmetric 
part of $K^{ab}_{\ \ i}$ remains in the field equations. Indeed, 
it is not hard to see that the remaining parts cancel already in 
the Lagrangian (1). Thus, the effective connection is given by 
$\tilde \Gamma^{ab}_{\ \ i} = \Gamma^{ab}_{\ \ i}(e) + \tilde K^{ab}_{\ \ i}$, 
where $\tilde K^{ab}_{\ \ i} = e_{ci} K^{[abc]}$.  Those  
differences 
between general relativity and Poincar\'e gauge theory are completely 
unrelated to the discussion of this paper, and  the  
use of the symbols 
$\nabla_i$ and $ \tilde \Gamma^{ab}_{\ \ i} $ is a convenient  way of 
treating both cases at the same time. 

For the study of the time evolution of $\psi$, the Schroedinger 
form of the Dirac equation, 
\begin{equation}
H \psi = i \partial_t \psi,  
\end{equation}
is especially adapted. We use greek indices $\alpha, \beta \dots $ 
to denote the spatial part of $a,b \dots $ (i.e., $ a = (0,\alpha)$), 
and indices $\mu, \nu \dots $ for the spatial part of the 
spacetime indices $i,j,k \dots $ and $t $ for the time component, 
(e.g., $m = (t, \mu)$). With the metric $g_{ik} = e_i^a e_k^b \eta_{ab}$ 
and using $\gamma^t \gamma^t = g^{tt}$, we find explicitely 
\begin{equation}
H = \frac{1}{g^{tt}} ( \gamma^t m - i \gamma^t \gamma^{\mu} \nabla_{\mu}) 
- \frac{1}{4} \tilde \Gamma^{ab}_{\ \ t} \sigma_{ab}. 
\end{equation}
In the flat limit ($\tilde \Gamma^{ab}_{\ \ i} = 0,\  e^a_i  = \delta^a_i$), 
this reduces to the well known expression $H = \gamma^0 m 
- i \gamma^0 \gamma^{\mu} \partial_{\mu} = \beta m + \vec \alpha \cdot \vec p$.

The Dirac Hamiltonian in curved spacetime has been used in the past, see, 
e.g., \cite{tiomno,audretsch,fischbach, casini, obukhov}, with the focus 
mainly on the coupling to gravity in the weak field limit. Although it 
 has been shown  that $H$ is not generally 
hermitian \cite{tiomno,audretsch}, 
this is not widely known, since in most relevant cases, the gravitational 
fields are static, and $H$ is hermitian. In this article, we will analyze
in detail the hermiticity properties of $H$.    

Before anything can be said about the hermiticity, we have to 
define a scalar product. One could try to construct this product 
such that certain conditions, like  covariance, or hermiticity of 
$H$ are a result of the  definition. However, there exists an even 
more physical requirement that leads straightforwardly to 
a consistent scalar product, namely the possibility of the normalization 
of the wave function. 
 
Indeed, using the Dirac equation (2) and its hermitian conjugate 
(\textit{hermitian conjugate} refers here only to Dirac space, i.e., 
the transposed complex conjugate equation), one derives the following 
conservation law 
\begin{equation}
\partial_i (e  \bar \psi \gamma^i \psi ) = 0, 
\end{equation}
where $e = \det e^a_i = \sqrt{-g}$. Integrating and omitting boundary terms, 
we get 
\begin{equation}
0 = \frac{\de }{\de t} \int \sqrt{-g}\ \bar \psi \gamma^t \psi\ \de^3 x
=  \frac{\de }{\de t} \int \sqrt{-g}\ \psi^{\dagger} (\gamma^0 
\gamma^t) \psi\ \de^3 x. 
\end{equation}
Therefore, obviously, the only way we can obtain a constant in time 
normalization is by using the inner product 
\begin{equation}
(\psi_1, \psi_2) = \int \sqrt{-g}\ \psi_1^{\dagger}(\gamma^0 \gamma^t) \psi_2 
\ \de^3 x, 
\end{equation}
or, in an explicitely covariant form 
\begin{equation}
(\psi_1, \psi_2) = \int \sqrt{-g}\ \bar \psi_1 \gamma^i  \psi_2\ 
\de S_i.  
\end{equation}
This product  is covariant and  
positive definite and the usual properties (linearity,  $(\psi_1,\psi_2) = 
(\psi_2,\psi_1)^*$) are easily verified. In the flat limit, where $\gamma^t = 
\gamma^0$, it reduces  to the special relativistic expression  
$(\psi_1,\psi_2) = 
\int \psi_1^{\dagger}\  \psi_2\ \de^3 x$. There is nothing else we could 
possibly require from an inner product. 

With (7) or (8), it can be directly checked whether $H$ from (6) is 
hermitian or not, i.e., whether we have $(\psi, H \psi ) = (H\psi, \psi)$
or not. This involves, however, quite a few partial integrations and a lot 
of $\gamma-$matrix shuffling. Fortunately, there is a much easier way 
to get to the answer.   

We start from  equation (6) and write 
\begin{displaymath}
 0 = \int\left[
 \sqrt{-g}\ [\partial_t \psi^{\dagger} \gamma^0 \gamma^t \psi 
+ \psi^{\dagger} \gamma^0 \gamma^t \partial_t \psi] 
+[(\partial_t \sqrt{-g}) \psi^{\dagger} \gamma^0 \gamma^t \psi 
+ \sqrt{-g} \psi^{\dagger} \gamma^0 (\partial_t \gamma^t) \psi]\ \right]
\de^3 x,  
\end{displaymath}
then use $\gamma^t_{,t} = e^t_{a,t}\gamma^a$, as well as the 
Dirac equation $H \psi = i \partial_t \psi$ and its conjugate $\psi^{\dagger}
H^{\dagger} = - i \partial_t \psi^{\dagger}$. (We remind once again that 
$H^{\dagger}$ is the hermitian conjugate with respect to Dirac space only, but 
we take the additional 
convention that its differentiation operators act to the left.) 
After some simple manipulations, we finally find 
\begin{equation}
 0 = i [(H\psi, \psi) - (\psi, H \psi)] + (\psi, 
[\partial_t  \ln \sqrt{-g\  g^{tt}}\ ] \psi), 
\end{equation}
showing that $H$ is, in general, not hermitian. In view of 
$g^{tt} \det g_{ik}  = \det g_{\mu \nu}$, this will be the case whenever 
the determinant of the three dimensional  space metric $g_{\mu \nu}$ 
is time dependent.  Moreover, we can define a hermitian operator 
\begin{equation}
\tilde H = H + \frac{i}{2} \partial_t \ln \sqrt{- g g^{tt}}, 
\end{equation}
satisfying 
\begin{equation}
(\tilde H \psi, \psi ) = (\psi, \tilde H \psi). 
\end{equation}

Thus, it turns out that the Hamiltonian in a gravitational background is 
not generally hermitian. We should
note, however, that in most cases, $\det g_{\mu \nu}$ will be 
time independent. This holds notably for static and stationary spacetimes, 
but also for gravitational waves when the usual gauge conditions are 
used in the linear approximation (see \cite{landau}). 
Counterexamples can be found, e.g.,  
in exact wave solutions of general relativity, as described in 
\cite{landau}, \S 109, and, most importantly, in cosmological 
solutions \cite{audretsch}.   

Let us also remark that, as we have claimed at the begin of the section, 
the final conclusion is  
completely independent of the eventual presence of torsion fields, since 
the non-hermitian term in (9) depends only on the spacetime metric.

\section{Field Hamiltonian} 

The main problem we face now is the interpretation of both $H$ and $\tilde H$. 
On one hand, we have a non-hermitian operator, with, in general, 
 complex eigenvalues, but with a straightforward interpretation as generator 
of the time evolution of $\psi$, see (3). On the other hand, the 
hermitian operator $\tilde H$, with real eigenvalues, was defined in a rather 
ad hoc manner, in order to absorb the non-hermitian terms found in (9). 

It turns out, however, that there is a straightforward, canonical, way 
to arrive at $\tilde H$. Indeed, the expectation value of $\tilde H$ 
is simply the field Hamiltonian  of the Dirac field, or in other words, 
the energy. 

In order to show this, we repeat some of the steps performed in
\cite{leclerc}. For convenience, we work in the Poincar\'e gauge 
theory framework, i.e., the Lagrangian (1) depends on the fields 
$\psi, \bar \psi$ and their derivatives, 
as well as on the gravitational background fields 
$\Gamma^{ab}_{\ \ i}, e^a_i$. Then, we consider the derivative 
of the Lagrangian density $\mathcal L = \sqrt{-g} L$, 
\begin{equation}
 \partial_i \mathcal L = \frac{\partial  \mathcal L }{\partial \psi}
\psi_{,i} + \frac{\partial \mathcal L }{\partial \psi_{,m}} \psi_{,m,i} 
+
\bar \psi_{,i} \frac{\partial  \mathcal L }{\partial \bar \psi}
 + \bar \psi_{,m,i} \frac{\partial \mathcal L }{\partial \psi_{,m}} 
 + \frac{\partial 
\mathcal L}{\partial \Gamma^{ab}_{\ \ m}} \Gamma^{ab}_{\ \ m,i} 
+ \frac{\partial \mathcal L}{\partial e^a_m} e^a_{m,i}. 
\end{equation}
Using the Dirac equations for $\psi $ and $\bar \psi$, we find 
\begin{equation}
[\sqrt{-g}\ t^i_{\ k}]_{,i}  = - \sigma_{ab}^{\ \ m} 
\Gamma^{ab}_{\ \ m,k} + T^m_{\ a} e^a_{m,k}, 
\end{equation}
where 
\begin{equation}
t^i_{\ k} = \frac{\partial L}{\partial (\partial_i\psi) } \ \partial_k \psi 
\ +\  \partial_k \bar \psi\ \frac{\partial L}{\partial (\partial_i \bar \psi)} 
\ -\  \delta^i_k L 
\end{equation}
is the canonical stress-energy tensor of the Dirac field, while 
$\sigma_{ab}^{\ \ m} = e^{-1} \delta \mathcal L / \delta \Gamma^{ab}_{\ \ m}$
and $T^m_{\ a} =- e^{-1} \delta \mathcal L / \delta  e^a_m $ 
are the spin density and the gravitational stress-energy tensor of 
the Dirac field (i.e., the sources of the gravitational field equations, 
see \cite{leclerc} for details). 

As outlined in \cite{leclerc}, the field momentum is given by 
\begin{equation}
\mathcal P_k = \int \sqrt{-g} \ t^i_{\ k}\ \de S_i = \int \sqrt{-g}\ 
t^t_{\ k}\ \de^3 x, 
\end{equation}
and the integrated version of (13) reads 
\begin{equation}
\frac{\de }{\de t} \mathcal P_k = - \int \sqrt{-g}\ [\sigma_{ab}^{\ \ m} 
\Gamma^{ab}_{\ \ m,k}- T^m_{\ a} e^a_{m,k}]\ \de^3 x. 
\end{equation}
This is the usual form of a canonical conservation law, stating that 
$\mathcal P_k $ (for some component $k$) is conserved whenever 
the background fields $e^a_m$ and $\Gamma^{ab}_{\ \ m}$ do not 
depend on the specific coordinate $x^k$. (Quite similarly, if there were, 
in addition, a 
background Maxwell field, there would appear at the r.h.s. of (16) a 
term $\int \sqrt{-g}\ j^m A_{m,k}\ \de^3 x$, 
with the electric current density $j^m$.) 

Especially, for the field Hamiltonian $\mathcal  H  = \mathcal P_t = 
\int \sqrt{-g}\ t^t_{\ t}\ \de^3 x$, we have 
\begin{equation}
\frac{\de }{\de t} \mathcal H = - \int \sqrt{-g}\ [\sigma_{ab}^{\ \ m} 
\Gamma^{ab}_{\ \ m,t}- T^m_{\ a} e^a_{m,t}]\ \de^3 x,  
\end{equation}
meaning that $\mathcal H$  is conserved whenever the background fields 
are time independent. These observations (and the fact that 
$t^i_{\ k}$ can be generalized to include the gravitational fields themselves 
in a way that $\mathcal P^{total}_k$ is conserved, see \cite{leclerc}) 
are sufficient to identify $\mathcal P_k$ as the physical momentum of the
system, and especially $\mathcal H$ as the field energy.  

We have already derived in \cite{leclerc} 
the explicit expression for $\mathcal H$ 
in the flat limit, 
it is the well known expression 
\begin{equation}
\mathcal H = \int \psi^{\dagger} H \psi\ \de^3 x = (\psi, H \psi). 
\end{equation}
Therefore, in  flat space, the field Hamiltonian (which plays, in 
quantum field theory, the role of the generator of time translations), 
is simply given by the expectation value of the Dirac Hamiltonian $H$, 
i.e., of the generator of the time translations of the 
quantum mechanical wave function.  

From $t^t_{\ t} = (i/2) (\bar \psi \gamma^t \partial_t \psi - 
\partial_t \psi \gamma^t \psi)$, partially integrating using (6) 
and  then the Dirac equation, or, alternatively, first the Dirac equation 
and then equation (9),  
it is not hard to show that in the general case, we find 
\begin{equation}
\mathcal H = \int \sqrt{-g}\  \psi^{\dagger}(\gamma^0 \gamma^t) 
 \tilde H \psi\ \de^3 x = (\psi, \tilde H \psi),  
\end{equation}
with $\tilde H $ from (10). 

This result not only provides us  with an interpretation of $\tilde H$, 
namely as  
the operator whose expectation values correspond to the energy of 
the system, but in addition, it justifies, a posteriori, the 
introduction of this arbitrarily defined  operator. 

Consequently, the hermitian $\tilde H$ 
should be interpreted as physical energy operator, while $H$ has simply  
the mathematical interpretation of the time evolution operator. This 
interpretation can be further founded on the following. If we 
define, for some operator $A$,  the operator $\de A / \de t$ 
in the conventional  way, i.e., by  requiring  
\begin{equation}
(\psi, \frac{\de A}{\de t} \psi) \equiv \frac{\de }{\de t} (\psi, A \psi), 
\end{equation}
then,  as is easy to show,  with the  inner product (7), 
we arrive at  the usual equation  
\begin{equation}
\frac{\de }{\de t} A = i [H,A] + \frac{\partial A}{\partial t}.
\end{equation}
This relation, together with (3), clarifies the meaning of the operator 
$H$. 

The fact that both operators do not coincide is not really a problem, 
since in practice, the use of energy eigenstates makes only sense  
 when the 
energy is conserved. This holds for time independent background 
fields, and in that case, we have $H = \tilde H $ anyway. 

However, it can easily be shown  that the momentum 
operator $p_{\mu} = i \partial_{\mu} $ is not hermitian, but 
that, similar to (9), we have the relation  
\begin{equation}
 0 = i [( p_{\mu} \psi, \psi) - (\psi, p_{\mu}  \psi)] + (\psi, 
[  \partial_{\mu}  \ln \sqrt{-g\  g^{tt}}\ ] \psi), 
\end{equation}
and consequently,  in analogy to (10), the hermitian operator is 
given by 
\begin{equation}
 \tilde  p_{\mu}  = p_{\mu} +\frac{i}{2} \partial_{\mu} 
\ln \sqrt{- g g^{tt}}.  
\end{equation}
Again, it turns out that the expectation value of this hermitian 
operator is identical to the field momentum $\mathcal P_{\mu}$ from 
(15), i.e., 
\begin{equation}
\mathcal  P_{\mu} = \int \sqrt{-g}\  \psi^{\dagger}(\gamma^0 \gamma^t) 
 \tilde p_{\mu} \psi\ \de^3 x = (\psi, \tilde p_{\mu} \psi),  
\end{equation}
in analogy to (19). In contrast to the situation for the 
Hamiltonian, the correction term in (23) will be different from zero 
 even for the most simple field configurations. 

Finally, though unrelated to our discussion, we would like to  remark 
that $T^a_{\ m}$ in (16) contains the full connection $\Gamma^{ab}_{\ \ i}$, 
and thus the full contortion tensor $K^{ab}_{\  \ i}$ (see \cite{hehl} 
for the explicit expression). Therefore, although only the axial 
 part of $K^{ab}_{\ \ i}$  couples to the  Dirac field 
in (1), the vector and tensor parts may  nevertheless influence 
the field momentum through (16).

\section{Non-unitary transformations}

In this section, we will make contact with the results previously 
obtained in literature. 

First, consider a transformation $A$ acting on $\psi$ and on operators 
$\mathcal O$ as  
\begin{equation}
\psi \rightarrow \psi' = A \psi, \ \ \mathcal O \rightarrow \mathcal O'
= A \mathcal O A^{-1}, 
\end{equation}
where, at the same time, the scalar product (7) goes over to the 
{\it flat} scalar product, i.e., 
\begin{equation}
(\psi_1, \psi_2) = 
 \int \sqrt{-g}\ \psi_1^{\dagger}(\gamma^0 \gamma^t) \psi_2 
\ \de^3 x 
\ \  \rightarrow\ \  <\psi_1, \psi_2> = \int \psi_1^{\dagger} \psi_2\ 
\de^3 x 
\end{equation}
We will refer to the tranformed quantities as being in the 
new representation. 
We require from $A$ that it 
does not change physical quantities, i.e., eigenvalues and expectation 
values. Therefore, we must have  
\begin{equation}
(\psi_1, \psi_2) = < \psi'_1, \psi'_2>, 
\end{equation}
which leads to the condition 
\begin{equation}
A^{\dagger} A = \sqrt{-g}\ \gamma^0 \gamma^t. 
\end{equation}
If the tetrad is in diagonal form, we have $\gamma^t = e^t_0 \gamma^0$, 
and (28) this can be satisfied by 
\begin{equation}
A = (-g\ g^{tt})^{1/4}. 
\end{equation}
If the tetrad is not diagonal, we first Lorentz rotate the vector $e^t_a$ such 
that $e^t_{\alpha} = 0$. (Then, of course, the gauge freedom 
is restricted, which is not unproblematic, because the Hamiltonian, as opposed 
to the Lagrangian, is not gauge invariant.) 

For $A$ satisfying (28), all expectation values remain 
unchanged, especially the expectation value of $H$. Also, it is clear that 
if an operator $\mathcal O$ is hermitian with respect to $(\ , \ )$, then, 
the transformed operator $\mathcal O'$ will be hermitian with respect 
to $<\ , \ >$. Therefore, (25) is a purely mathematical operation which 
does not affect the physics, just as are unitary transformations in 
special relativity. (In our case, the non-unitary part of $A$ is 
absorbed by the simultaneous  change of the scalar product.) 

The Dirac equation takes the form 
\begin{eqnarray}
H' \psi' &=& AHA^{-1}A \psi = A i \partial_t \psi \nonumber \\
&=& i \partial_t \psi' - i(\partial_t A) A^{-1} \psi', 
\end{eqnarray}
or, with the help of (29), 
\begin{equation}
i \partial_t \psi' = [H' + \frac{i}{2} \partial_t \ln \sqrt{-g\ g^{tt}}] 
\psi'. 
\end{equation}
If we compare this with (10), we see that in the new representation, 
the operator $i \partial_t$  can be identified with the operator $\tilde H$, 
i.e., we have $i \partial_t = \tilde H'$. Clearly, just as $\tilde H$
was hermitian w.r.t. the original scalar product, 
$i \partial_t $ is hermitian w.r.t. the flat scalar product. Thus, in 
the new representation, the energy operator is given directly by the time 
evolution operator. 

Note that the operator $H'$, as before, is not hermitian, it satisfies 
equ. (9), where $(\ ,\ )$ is replaced by $<\ ,\ >$.  
 
On the other hand, it has been shown that by the transformation 
\begin{equation}
H'' = A H A^{-1} - i A \partial_t A^{-1}, \ \ \psi''= A \psi, 
\end{equation}
where $A$ still satisfies (28), 
the Hamiltonian $H$ is brought into a form that is hermitian w.r.t. 
the flat scalar product (see  \cite{tiomno,audretsch}). It is not 
hard to show that this is indeed the case, but we wish to point out 
that there is more involved here then a mere transformation. 

First of all, it is clear that $H''$ does not have the same eigenvalues 
as $H$. Thus, both operators are physically different. Secondly, as 
opposed to (25), which is applied to every operator, the transformation 
(32) acts specifically on the Hamiltonian. In the same way, 
you can bring, e.g., the momentum operator $p_{\mu}  = i\partial_{\mu}$ 
into the hermitian form 
\begin{equation}
p_{\mu}'' = A p_{\mu} A^{-1} - i A \partial_{\mu} A^{-1}, 
\end{equation}
which is clearly different from (32). Therefore, such transformations 
are not merely mathematical tools, but are physical redefinitions 
of the operators $H$ and $p_{\mu}$. This becomes even more clear if 
we split the transformations into two steps. First, we perform the 
mathematical transformation (25), acting on all operators, but 
not changing any physical quantities. Then, in a second step, 
we {\it define} $H'' = H' -i  A \partial_t A^{-1}$ (and similarly, 
$p_{\mu}'' = p' - i A \partial_{\mu} A^{-1}$), which is simply the 
definition of new, hermitian operators.      

It is now clear that in the previous  sections, 
we did exactly the same thing, namely we redefined $H$ and $p_{\mu}$ in 
order to find hermitian operators (see (10) and (23)), with the only 
difference that we did it directly in the initial form, i.e., without 
performing the first step, which is completely unrelated to the 
hermiticity properties.  

To conclude, we showed that, apart from a purely mathematical 
change of representation (25), whose purpose is to bring  the 
scalar product into a flat form (such that the hermiticity properties 
of the operators can be directly read off), the {\it transformations}
(32) and (33)  are exactly equivalent to the redefinitions (10) and (23). 
Therefore, the interpretation of (32) and (33) is  that they are 
physically non-trivial redefinitions and not simply mathematical 
transformations. 
They  are just as arbitrary as our equations (10) and (23), i.e., they 
 can only 
be justified by physical arguments, especially  
by the fact that the resulting operators are hermitian. (Unfortunately, 
the transformation of the form (32) is sometimes confused with a 
change of representation, and said not to change physical quantities, 
because it leaves the Dirac equation $H \psi 
= i \partial_t \psi $   form-invariant. However, this does not 
change the fact that $H''$ is a different operator from $H$, as can 
be seen from the eigenvalues. This is because, with a change of 
the representation, the operator $i \partial_t $ is transformed too.) 

The result of section 2, which relates the new operators to the 
field energy and the field  momentum is therefore an a posteriori
justification of the transformations (32) and  (33) showing that   
the new, hermitian operators, really correspond to the physical 
quantities they are supposed to.

\section*{Acknowledgments}

This work has been supported by EPEAEK II in the framework of ``PYTHAGORAS 
II - SUPPORT OF RESEARCH GROUPS IN UNIVERSITIES'' (funding: 75\% ESF - 25\% 
National Funds). Special thanks to Y.N. Obukhov for suggesting a possible 
relation 
of our result to the non-unitary transformations previously used in 
literature and for pointing out important references.

\end{document}